# Addendum to: "Kolmogorov complexity of sequences of random numbers generated in Bell's experiments" (series of outcomes).

Marcelo G. Kovalsky, Alejandro A. Hnilo and Mónica B. Agüero

*CEILAP, Centro de Investigaciones en Láseres y Aplicaciones, UNIDEF (MINDEF-CONICET);*
*CITEDEF, J.B. de La Salle 4397, (1603) Villa Martelli, Argentina.*
email: ahnilo@citedef.gob.ar

December 13th, 2018.

In the mentioned paper we presented results of the estimation of Kolmogorov complexity of sequences of random numbers generated in a famous Bell's experiment, aimed to study the security of QKD. We focused on series of time differences between successive detections of coincidences, and found that randomness cannot be taken for granted. It was then criticized that the theorems that demonstrate the randomness of series produced in Bell's experiments involve series of measurement *outcomes*, not of measurement *times*. Here we reply to this objection and present data of series of outcomes, showing that the conclusions in the main paper are valid also in this case.

In a recently published paper [1] we analyzed the randomness of series formed by the time elapsed between successive coincidences in the Innsbruck experiment [2]. We found that the series cannot be considered random in 5 over 21 (set of best runs). Besides, in one of the runs (named *longtime*) it was possible to reconstruct an attractor in phase space and predict values of the series in a QKD scheme. We concluded that it is not safe in the practice taking the randomness of quantum-measurement-produced series *for granted*. We presented these results in several conferences during this year, and received the objection that the theorems ensuring randomness of quantum measurements [3,4] apply to measurement *outcomes*, not to measurement *times*. Although we briefly considered this issue in [1] the controversy remains, so we think pertinent to present additional data regarding the randomness of series of *outcomes*, to show that the conclusions reached for series of *times* are valid in this case too. The present contribution is not intended to be self-contained but an addendum to [1], so that paper must be at hand in what follows.

Let review in few words why we had chosen to study the randomness of times instead of outcomes in the first place:
*i)* The randomness of series of outcomes had been previously studied [5,6], and found to be surprisingly poor. In [6] (where the aim was to get a reliable source of random numbers) the cause of this drawback was identified in the blind time of the detectors, and successfully solved by using the *time between coincidences* (above or below the average) to generate a Borel-normal binary string. It was then natural to focus on this type of series to study their Kolmogorov complexity $K$. In other words: series of outcomes were already known to be not always random.
*ii)* Series of times between successive coincidences are longer (= better statistics). The results are hence more reliable.
*iii)* The series of times may allow predicting the series of outcomes, as it was demonstrated in the case of the run *longtime*, at least in a QKD scheme.

In what follows, we present results of the calculation of randomness of series of *outcomes*. As this contribution is intended to be a short addendum, we do not present an exhaustive study of all available data (as we did in [1]) but just a set of representative examples of series of outcomes, namely: a run which series of time coincidences was found random with complexity $K \approx 1$ (*longdist0*); a run that was found random and $K >> 1$ (*longdist1*) and a run that was found not random (*longdist35*, $K=0.34$). We include in this set, as a reference, a series of outcomes of randomness certified by the impossibility of superluminal signals [7], obtained in a recent loophole-free Bell's experiment [8] and which is available in the web [9]. Recall the Innsbruck experiment [2] closed the locality or predictability loophole, but it was not detection-loophole-free (it was not intended to).

In [1], a series was considered random if it had $K>0.9$ and passed the first 6 tests of the NIST battery [10]. Here we use the complete set of 15 tests in the battery. None of the series mentioned above passes test #9 ("Maurer universal statistical test"), but this failure is to be expected, because this test is reliable for series of length $N > 3.88 \times 10^5$. Here we call a series "random" if it passes the full battery of tests (excepting #9) and has $K \approx 1$. We recall that the value of the complexity can be only estimated. Here we use the realization of Lempel and Ziv algorithm developed in [11] and implemented in [12].

The series of outcomes in the Innsbruck experiment are in the second column in Fig.1 of [1], coded with the numbers {0,1,2,3} to indicate the analyzer setting (equivalently, the voltage applied to the modulator) and the detector that fired. The number "0"("2") means that detector "0"("1") fired, with no voltage applied to the modulator. The number "1"("3") means that detector "0"("1") fired, with voltage applied to the modulator. In this way, from each run, 4 series of random numbers are extracted: two stations (Alice and Bob), and one series for each of the two analyzers' settings. The detector that fired, for a given setting, determines a 1 or a 0 in the series. Recall that only detections that produce *coincidences* between the two stations are taken into account. The results are summarized in the Table 1.

The first visible result is that all series have $K \approx 1$ and can be considered algorithmically random, as it was mentioned in [1]. When the first 6 tests of NIST are applied, only 7 of the 15 series can be considered statistically random. The proportion of non-random outcome series (≈½) is hence larger than for the time series (≈¼). This result is consistent with the ones reported in [6]. If the full battery of NIST tests is applied, only 4 series survive. One of them belongs to a run that was random with $K>>1$ (*longdist1*, Bob station, HV applied), another one to a run that was *not* random (*longdist35*, Alice station, no HV applied). These examples show that the randomness of the outcomes' sub-series has no evident relationship with the randomness of the complete series. With the same purpose, we show the results for the whole column on the right of Fig.1 in [1] (that is, without separating it into 4 sub-series) for run *longdist35*. Both series are algorithmically random, but Alice's series fails to pass 8 of 15 tests of statistical randomness (including 4 of the first 6). Perhaps surprisingly, Bob's series is one of the "fully" random ones. Finally, the quantum-certified random series obtained in the loophole-free setup is random, as expected.

In summary: we present results verifying that series of *outcomes* obtained in quantum measurements are not always random in the practice. In fact, they are random less often than series of *times* are. Hence, we confirm the results reported in [5,6] and also our main conclusions in [1]. The reason why deviations from expected randomness occur is beyond reach, because the Innsbruck experiment was dismantled long ago. It may be argued that the loophole-free generated series passes all our tests of randomness, so that the cause may be in the detection loophole (which was not ruled out in the Innsbruck's experiment). Yet, in our opinion, a study on a larger set of data is necessary before something can be said about a hypothetical influence of that loophole.

**Acknowledgments.**

Many thanks to A.Khrennikov and J.G.Hirsch for the encouragement, observations and advices, and to D.Mihailovic for his help. Many thanks again to G.Weihs for giving us the complete files of the Innsbruck experiment, and to L.K.Shalm and his group for allowing free access to the loophole-free quantum-certified random file. Thanks also to C.Shmiegelow, I.García Mata, F.Toscano and R.Roy for discussions during several conferences and talks, which triggered the redaction of this Addendum. This work received support from the grants: "Search for vulnerabilities of fundamental origin in QKD" N62909-18-1-2021 Office of Naval Research Global (USA), and "Desarrollo de láseres de estado totalmente sólido y de algunas aplicaciones" PIP 2017 0100027C CONICET (Argentina).

**References.**

| Series of outcomes. | Complexity | NIST (RND=?) | $S_{CHSH}$ | $N$ |
|---|---|---|---|---|
| **Longdist0, Alice, setting = 0** | 1.017 | NO | 2.53 | 9676 |
| **Longdist0, Alice, setting = 1** | 1.030 | NO | 2.53 | 10302 |
| **Longdist0, Bob, setting = 0** | 1.023 | NO | 2.53 | 9893 |
| **Longdist0, Bob, setting=1** | 1.023 | NO | 2.53 | 10085 |
| **Longdist1, Alice, setting = 0** | 1.025 | NO | 2.63 | 10848 |
| **Longdist1, Alice, setting = 1** | 1.027 | yes (no) | 2.63 | 9859 |
| **Longdist1, Bob, setting = 0** | 1.018 | NO | 2.63 | 10043 |
| **Longdist1, Bob, setting = 1** | 1.016 | yes | 2.63 | 10664 |
| **Longdist35, Alice, setting = 0** | 1.032 | yes | 2.73 | 8638 |
| **Longdist35, Alice, setting = 1** | 1.016 | NO | 2.73 | 6365 |
| **Longdist35, Bob, setting =0** | 1.033 | yes (no) | 2.73 | 7741 |
| **Longdist35, Bob, setting = 1** | 1.023 | yes (no) | 2.73 | 7262 |
| **Longdist35, all Alice outcomes** | 1.021 | NO | 2.73 | 15003 |
| **Longdist35, all Bob outcomes** | 1.026 | yes | 2.73 | 15003 |
| **Bierhorst et al.** | 1.044 | yes | $J = 1.41 \times 10^{-5}$ | 1024 |

TABLE 1: Summary of results. The third column is "NO" if the run does not pass one of the first 6 test of NIST battery (this is to facilitate comparison with results in [1]). It is "yes (no)" if it passes the first 6 tests but fails to pass one of the remaining tests in the battery (excepting #9, see the main text). All series belong to the Innsbruck experiment [2] excepting the last one, which belongs to the loophole-free experiment [7-9]. The fourth column indicates the violation of the corresponding Bell's inequality. Note the last setup uses Eberhardt's inequality ($J \leq 0$) instead of CHSH ($S_{CHSH} \leq 2$). The last column is the length of each series.